\RequirePackage{fix-cm}
\documentclass[numbook,smallextended]{svjour3}       

\smartqed  
\usepackage{ifpdf}
 \ifpdf
\usepackage[pdftex]{graphicx}
\newcommand{\pix}{jpg}
 \else
\usepackage[dvips]{graphicx}
\newcommand{\pix}{jpg}
 \fi

\usepackage{color}

\usepackage{url}
\usepackage{amsmath}
\usepackage{amssymb}
\usepackage{mathptmx}      

\spnewtheorem{defi}[subsection]{Definition}{\bfseries}{\rmfamily}
\def\makeheadbox{{%
\hbox to0pt{\vbox{\baselineskip=10dd\hrule\hbox
to\hsize{\vrule\kern3pt\vbox{\kern3pt
\hbox{\bfseries
Extended version of a short paper
with the same title 
presented on 4/4 2019,
}
\hbox{\bfseries
at WCC'2019 in Abbaye de Saint-Jacut de la Mer, France,
31/3 - 5/4 2019.
}
\kern3pt}\hfil\kern3pt\vrule}\hrule}%
\hss}}}
\journalname{
Extended version of a short paper
presented at WCC'2019, April 2019.
}
\begin{document}

\def\mytitl{Invariant Hopping Attacks on Block Ciphers}
\title{\mytitl}


\author{Nicolas T. Courtois
}

\institute{
Nicolas T. Courtois, ORCID=0000-0003-0736-431X
\at University College London, Gower Street, London, UK \\
              \email{n.courtois@cs.ucl.ac.uk}           
}

\date{
\\[-17pt] 
University College London, UK}

\maketitle

\begin{abstract}
Block ciphers are in widespread use since the 1970s.
Their iterated structure is prone to numerous round invariant attacks
for example in Linear Cryptanalysis (LC).
The next step is to look at non-linear polynomial invariants cf. Eurocrypt'95.
Until recently, researchers have found extremely few such attacks,
with some impossibility results \cite{BeiCantResNL,GenNonLinRC,FiliolNotVuln}.
Eventually recent papers show how to construct polynomial invariant attacks for block ciphers,
however many such results were of degree 2, cf. \cite{BLC,TodoNL18,BackdTut}.
In this paper we propose a new incremental methodology
for constructing high degree polynomial invariant attacks
on block ciphers.
A trivial attack on one cipher setup
will be transposed to show the existence
of a more advanced attack on a
stronger cipher in several steps.
The key tool is
the manipulation of the roots of the so called Fundamental Equation.
Examples are constructed with an old historical block cipher T-310.

\end{abstract}

\noindent
{\bf Acknowledgements.} 
Special thanks to Felix Ulmer for many inspiring discussions
we had during WCC'2019,
which enabled me to substantially
improve my work on cryptanalysis of T-310.
We also thank all the anonymous referees for reading our paper very carefully
and spotting several mistakes.
We thank UCL students and our research assistant Matteo Abbondati
for helping me to develop and validate some examples of attacks which are listed here.
We thank Jacques Patarin, Gilles Macario-Rat, Jean-Jacques Quisquater
and Joseph Pieprzyk for interesting discussions.


%

\vskip-3pt
\vskip-3pt
\section{Introduction, Non-Linear Cryptanalysis}
\label{intro}
\vskip-3pt

The concept of cryptanalysis with non-linear polynomials
a.k.a. Generalized Linear Cryptanalysis (GLC)
was introduced
at Eurocrypt'95, cf. \cite{GenLinear1}.
A key question is the existence of round-invariant I/O sums: 
when a value of a certain polynomial is preserved
after 1 round.
Such properties are notoriously hard to find \cite{TodoNL18,BackdTut}.
There are $2^{2^n}$ possible invariants and 
systematic exploration is not feasible \cite{BeiCantResNL,BeiCantRevis,GenLinear2,GenNonLinRC}.
In this paper and unlike in \cite{TodoNL18} we focus on invariants which work for 100 $\%$
of the keys and we focus on stronger invariants which hold with probability 1 for one round.
In addition we look at an expensive historical and government
cipher 
where encryption is substantially more costly than in modern ciphers such as 3DES, AES, cf. \cite{FeistCommunist}.
The number of rounds used to encrypt each character is simply very large and most
standard cryptanalytic attacks will simply do not work,
or will be very far from being practically relevant.
However all this complexity is not that useful if we can discover invariant attacks working for any number of rounds.

This paper is organized as follows.
In Section 2 we explain the philosophy of the attacks and relation to mathematical
theories of invariants and group theory.
In Section 3 we introduce the question on polynomial invariants on ciphers and some useful notations.
In Section \ref{subsec2:constructive} we look at the question of Feistel ciphers
vs. general block ciphers.
In Section \ref{subsec2:spec} we
provide the specification of our block cipher T-310.
In Section \ref{sec:3} we explain that the problem of finding a one-round invariant
can be formalized as the problem of solving the so called Fundamental Equation $\mathrm{(FE)}$.
In Section \ref{sec:KT1Quadratic} we show several examples of
polynomial invariant attacks,
some of which are simple linear attacks.
We show how a linear invariant $\cal L$ on one cipher 
may hide the existence of another non-linear invariant property $\cal P$ on the same cipher. 
We do not stop here: in Section \ref{subsec4:degree4} we show that an invariant of degree 4 also exists,
and in further sections the simpler invariants will be removed and we keep ONLY one invariant of degree 4.
Then in Section \ref{SimpleInvP20Cycle9BiasedFEHomQuadBakdPaper551DifferentOpticHigherDegree4Better}
we modify the wiring in order to accommodate a more complex Boolean function. 
Then in Section \ref{subsec5:94} we show that two similar invariants may exist in one single cipher setup.
Finally in Section \ref{subsec5:95} we will remove all invariants of degree 4
and we will be left with a one single invariant of degree 8.
In few steps a pathologically weak cipher
becomes a substantially stronger cipher
and the attack becomes less obvious
and harder to discover.
In Section \ref{subsec5:96} and in Conclusion section we
explain what was achieved so far and present some open problems.

\vskip-9pt
\vskip-9pt
\section{Polynomial Invariant Attacks}
\label{sec:2}
\vskip-2pt

\subsection{Invariant Attacks, Partitioning Cryptanalysis, Linear Subspace Attacks}

\label{subsec1:Parti}
A general approach in
cryptanalysis 
considers arbitrary subsets of binary vector spaces.
This is called Partitioning Cryptanalysis (PC), cf. \cite{BannierPartBack,
HarpMassThm}. 
Our work is then a special case:
we study only partitions defined by the value (0/1) for a single Boolean polynomial, cf. \cite{BackdTut}.
This is less general but
properties are more intelligible
and follow clear rules of formal algebra.
A serious theory is nowadays being developed around what is possible or not to
achieve in partitioning and invariant attacks,
cf. \cite{BeiCantResNL,GenNonLinRC,FiliolNotVuln}.

In recent research there exist two major types of invariant attacks:
linear sub-space invariants \cite{Zorro,PrintSubspace,BeiCantResNL,FiliolNotVuln},
and proper non-linear polynomial invariants \cite{TodoNL18,BackdTut}, 
which are somewhat more general. 
In many cases (but not always) two types of attacks are the same:
an attack where the invariant is a product of simple linear polynomials $A,B,C,D$,
for example ${\cal P}=ABCD$ in Section \ref{subsec4:degree4}, is also simultaneously a linear subspace attack.
However the most general attack is an attack with a sum of products, for example
${\cal P}=AC+BD$
in Section \ref{subsec4:decomposeACBD} and it is irreducible so it cannot be expressed as a single product,
and therefore it is {\bf not} an attack
with a linear sub-space invariant \cite{BannierPartBack,BeiCantResNL,GenNonLinRC,Beyne18}.
This demonstrates that the product attack is NOT necessarily the most powerful attack. In general we work in polynomial rings
and both addition and multiplication are allowed and the polynomial
${\cal P}=AC+BD$ in Section \ref{subsec4:decomposeACBD} was verified to be irreducible.

\vskip-2pt
\vskip-2pt
\subsection{Invariants vs. Mathematical Theory of [Polynomial] Invariants}
\label{subsec1:introInvMaths}

A key observation
is that
a set of all possible invariants
for any block cipher is in general a ring:
a sum of two invariants is an invariant.
Likewise, a product of two invariants gives also another invariant.
We have a potentially rich mathematical structure, a ring.
However in many cases this ring seems\footnote{
This however we are not sure about, in most cases.
Additional invariants not anticipated by any given construction or method such as those studied
in this paper, are sometimes discovered, and many more invariants
at higher degree and large complexity could exist,
without being efficiently computable
or efficiently detectable by the attacker.
}
 to have only a small number of elements.
 One of the reasons for this is that when we multiply various
 Boolean functions frequently we get zero (annihilation) or one of the operands
 (absorption). This also means that it makes a lot of sense to study these rings
 as partially ordered set (or POSETs) with the partial order defined by
 the usual division of (Boolean) polynomials.
 A standard method to represent POSETs are Hasse diagrams, cf.
 Fig. \ref{LatticeWalk} below.

There exists an extensive theory of multivariate polynomial algebraic invariants 
\cite{
CrillyInvariantsHist} which
historically, in mathematics,
has studied mostly invariants w.r.t. linear transformations and
has rarely studied invariants with more than 5 variables
and in finite fields of small size.
In our work we study invariants w.r.t {\bf non-linear} transformations and 36
or more
variables
 over $GF(2)$.
More ample explanations on how the present work is simultaneously
very closely related and yet differs very substantially from questions typically
studied by mathematicians in the classical [Hilbertian]
invariant theory can be found in Section 2.4 in \cite{LackUnique}.
We summarize the main comparison points here. In classical work in maths we have:

\vskip-8pt
\vskip-8pt
\begin{enumerate}
\item [1)] invariants are polynomials of small degree,
\item [2)] they have only 2 sometimes up to 5 variables,
\item [3)] polynomials are over large fields and rings,
frequently algebraically closed or infinite (or both), or in fields with large characteristic,
\item [4)] invariants should not change when we operate a {\bf LINEAR} input variable transformation $L$,
a very important limitation,
\item [3+4]
makes that there is a scaling scalar or factor $\sigma$ in most invariants known in classical mathematics:
a determinant of the linear transformation $L$,
\item [5)] these invariants are in general multivariate polynomials.
\end{enumerate}
\vskip-3pt
\vskip-3pt

The common points with our work on block ciphers are (1) and (5)
%
and there are {\bf very substantial} differences as follows:

\vskip-7pt
\vskip-7pt
\begin{enumerate}
\item [2')] we work with many more variables for ${\cal P}$, typically between 8 and 36 at a time.
\item [3')] we work with $GF(2)$ mainly, which seems to be the best choice,
\item [4')] we work with invariants which remain the same after applying an extremely complex
{\bf non-linear} transformation sometimes called $\phi$,
or any power of it $\phi^k$
with extra variables which are secret (such as key bits)
or public (such as round constants or IV bits).
\item [3+4]
Here the scaling factor $\delta$ could only be equal to $1$ and should be omitted.
\end{enumerate}
\vskip-4pt

%

\vskip-9pt
\vskip-9pt
\subsection{A Group Theoretic Interpretation}
\label{subsec1:Group}

The question of invariants is very closely related and also to
the study of the groups generated by various cipher
transformations
This question was studied for a long time, since early 1970 in both Eastern Bloc
\cite{T-310An80,DESPrimitiveWernsdorf,AESWernsdorf}
and in the West,
\cite{KennyImprimitive,SalaGOSTgroup,invglc,WhiteningParadox}.

\vskip-7pt
\vskip-7pt
\begin{figure}[ht!]
\vskip-5pt
\hskip-10pt
\hskip-10pt
\begin{center}
\hskip-10pt
\hskip-10pt
\vskip-3pt
\vskip-3pt
\vskip-3pt
\includegraphics*[width=5.1in,height=3.2in]{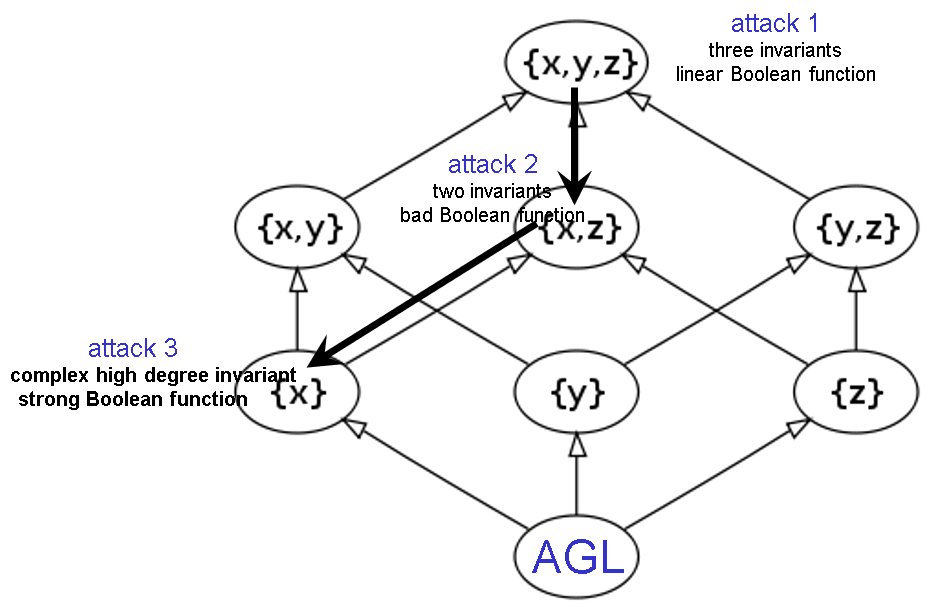}
\end{center}
\vskip-2pt
\vskip-2pt
\caption{Interpretation of our work in terms of walks in group lattices}
\label{LatticeWalk}
\end{figure}
\vskip-2pt
\vskip-2pt

It is very difficult to know the actual size of a group generated by cryptographic
transformations except that cipher designers have worked
very hard for decades on how to make it
extremely large and extremely complex.
What is however accessible to us is to see that certain groups are smaller
because each transformation satisfies a certain invariant.
In general different groups generated by different versions of a cipher
will be ordered by inclusion and we get a partially ordered set of groups.
More generally these groups can be embedded in a larger lattice of groups
which do not need to correspond to a particular version of a cipher,
but also characterised by additional (technical) conditions.
Then the whole incremental approach of invariant hopping which we study in this paper
can be interpreted as walks in group lattices.
This is illustrated on Fig. \ref{LatticeWalk}.

In this paper we will not study this approach in detail.
An open problem is to revisit known result on groups generated by block cipher
transformations and at the same time compute the full invariant ring for each group.
This is quite difficult because currently no efficient method
for computing the full invariant ring is known.

\vskip-3pt
\vskip-3pt
\subsection{Finding Advanced Invariant Attacks}
\label{subsec1:Discover}
The primary method
proposed in 2018 is through solving the so called Fundamental Equation or $\mathrm{FE}$ cf. \cite{BackdTut}.
Solving such equation(s),
or several such equations simultaneously,
{\bf guarantees} that we obtain a Boolean function for our cipher
and the polynomial invariant ${\cal P}$
which propagates for any number of rounds.
However
nothing guarantees that the $\mathrm{FE}$ equation has any solutions whatsoever.
%
In this paper we construct polynomial invariants
explicitly by modifying
something which worked.
A trivial attack on a weak cipher will be
transformed into a better or higher degree invariant attack on a stronger cipher
in several steps. We call this method ``invariant hopping'' illustrated
in Fig. \ref{LatticeWalk}.
The discovery of invariant properties by modification of other invariant properties
and learning from examples
is the main topic of this paper.


\vskip-9pt
\vskip-9pt
\subsection{The Question of Weak Keys and Backdoors}
\label{subsec1:WeakKeysBackd}

There are numerous constructions of weak ciphers in cryptographic literature,
cf. for example work related to the AES S-box \cite{invglc,WhiteningParadox},
and \cite{BannierPartBack,FiliolNotVuln} including work on choice of constants in
modern hash functions \cite{MaliciousSHA,MaliciousKeccak} which are also based on block ciphers.
In this paper, a weak key or rather a weak long-term key setup is primarily a tool to find some attack more easily,
and the same attack may be transposed to hold also for another (stronger) cipher setup.
In terms of backdoors an interesting question is how many entries in the truth table of an arbitrary Boolean function
need to be altered in order to make a given polynomial invariant attack work.
Recent work suggests that is some cases this number can be quite low, cf. \cite{BackdAnn}.
This is also what we will demonstrate in Section \ref{subsec5:95}.

\newpage
\vskip-9pt
\vskip-9pt
\section{Polynomial Invariants and Block Ciphers}
\label{sec:2}
\label{DefInvariant}
\vskip-2pt

We call ${\cal P}$ a polynomial invariant if the value of ${\cal P}$
is preserved after one round of encryption,
i.e. if

\vskip-7pt
\vskip-7pt
$$
{\cal P}(\mbox{Inputs})={\cal P}(\mbox{Outputs}).
$$
\vskip-2pt

Another notation is to write that

\vskip-7pt
\vskip-7pt
$$
{\cal P}= {\cal P}^{\phi}
$$
\vskip-2pt

where ${\cal P}^{\phi}$  denotes how our polynomial is transformed by our block cipher,
where one round of our block cipher is denoted by $\phi()$,
and
and we will later
define

\vskip-3pt
\vskip-3pt
$$
{\cal P}^{\phi} \stackrel{def}{=} {\cal P} \circ \phi =
{\cal P}(\phi(\mbox{Inputs})) = {\cal P}(\mbox{Transformed Outputs})
$$
\vskip-2pt
where $\circ$ is the composition of functions which by convention is applied (in order) from right to left.
However this notation is somewhat ambiguous.
On one side it is possible that it is not well defined because
one round of encryption $\phi$ depends on some extra parameters such as key or IV bits.
On the other side, the polynomial ${\cal P}$ is frequently very carefully chosen by the attacker so that
this would not be the case and numerous variables are already eliminated.

This works for any block cipher except that such attacks are notoriously hard to find
\cite{BeiCantResNL} in the last 20 years since  \cite{GenLinear1}.
In this paper we are going to work with one specific block cipher with 36-bit\footnote{
Block size could be increased and our attacks and methods would work all the same.} blocks.
The main point is that any block cipher round translates into relatively simple Boolean polynomials,
if we look at just one round.
We follow the methodology of \cite{BackdTut} in order
to specify the exact mathematical constraint, known as the Fundamental Equation or $\mathrm{FE}$,
so that we could have a polynomial invariant attack on our cipher.
Such an attack will propagate for any number of rounds (if independent of key and other bits).
In addition it makes sense following \cite{BackdTut} to consider
that the Boolean function inside the cipher is an unknown. 
We denote this function by a special variable $Z$.
We then see that our attack works if and only if $Z$ is a solution to
a certain algebraic equation [with additional variables].
The main interest of making $Z$ a variable is to find some strong attacks
in the cases where the Boolean function is extremely weak case, (e.g. $Z$ is linear)
and transpose them to stronger ciphers where $Z$ will be increasingly complex.

\noindent
We discard the attacks when the $\mathrm{FE}$ reduces to 0 a nd which work for any $Z$.
It appears that such attacks are quite rare cf. \cite{BackdTut}
and our later specific trick or method
for finding new attacks by manipulating the roots of $\mathrm{FE}$ would not work.

\vskip-2pt
\vskip-2pt
\subsection{Notation and Methodology}
\label{subsec2:notameto}
\vskip-2pt

In this paper the sign + denotes addition modulo 2, and frequently
we omit the sign * 
in products.
For the sake of compact notation we frequently use short or single letter variable names.
For example let $x_{1},\ldots, x_{36}$ be inputs of a block cipher each being $\in\{0,1\}$.
We will avoid this notation and name them with small letters $a-z$ and letters $M-V$ when we run out of lowercase letters.
We follow the backwards numbering convention of \cite{BackdTut} with $a=x_{36}$ till $z=x_{11}$
and then we use specific capital letters $M=x_{10}$ till $V=x_{1}$.
This avoids some ``special'' capital letters
following notations used since the 1970s \cite{FeistCommunist,T-310,T-310An80}.
We consider that each round of encryption is identical except that
they can differ only in some ``public'' bits called $F$ (and known to the attacker)
and some ``secret'' bits called $K$ also known as $S_1$, or $L$ also known as $S_2$.
Even though these bits ARE different in different rounds we will omit to specify in which round
we take them because our work is about constructing {\bf one round} invariants (extending to any number of rounds).
This framework covers most block ciphers ever made  
except that some ciphers would have more ``secret'' or ``public'' bits in one round.
The capital letter $Z$ is a placeholder for substitution of the following kind

\vskip-3pt
\vskip-3pt
$$
Z(e_1,e_2,e_3,e_4,e_5,e_6)
$$
\vskip-2pt

where $e_1\ldots e_6$ will be some 6 of the other variables.
In practice, the $e_i$ will represent a specific subset of variables
of type $a$-$z$, or other such as $L$.
Later $Z$ (and maybe another letter like $W$) needs to be replaced by a formula like:

\vskip-7pt
\vskip-7pt
$$
Z \leftarrow Z00+Z01*L+Z02*c+Z03*Lc+\ldots +Z62*cklfh+Z63*Lcklfh
$$
\vskip-2pt

where $Zij$ are coefficients of the Algebraic Normal Form (ANF).

\vskip-2pt
\vskip-2pt
\subsubsection{Polynomial Invariants}
\label{subsec2:polyinva}
\label{FirstGlimpseOnPSymmetries}
We are looking for arbitrary Boolean polynomial invariants.
%
For example say 
%
\vskip-4pt
\vskip-4pt
$$
{\cal P}(a,b,\ldots) = 
abc+abd+acd+bcd +
\ldots
$$
\vskip-2pt
could be an invariant after 1 round.
In this space some solutions are considered as trivial and are easy to find,
for example when ${\cal P}$ is a simple product of linear invariants cf. Appendix of \cite{BackdTut}.
For this reason some early papers on this topic emphasised
complex irreducible polynomials which were considered both less trivial
and harder to discover.
One of the main points in discovery of innovative attacks on block ciphers
is that non-trivial attacks with irreducible polynomial invariants of degree higher than 2 are
actually at all possible, and how to construct some (rather than just discover accidentally).
In fact we do NOT and should not restrict our attention to irreducible polynomials.
A fundamental point is that even when ${\cal P}$ is just a simple product of linear invariants,
it is possible that the linear invariants will not work and only their product works.
Or that we can subsequently remove the simpler invariants (!).
Several examples of this are shown in this paper.



\vskip-2pt
\vskip-2pt
\section{A Constructive Approach Given the Cipher Wiring}
\label{subsec2:constructive}
\vskip-3pt

We consider an arbitrary block cipher\footnote{It could also be a stream cipher
where the state is transformed in a bijective way in the same way
as in an unbalanced Feistel cipher,
or a stream cipher based on a block cipher.}.
Our attack methodology starts\footnote{
Our approach is to find invariant attack starting from arbitrary rounds ANFs
is at the antipodes compared to \cite{invglc,WhiteningParadox} where the ciphers are very special.
}
 from a given block cipher specified by its ANFs for one round.
Specific examples are shown for T-310, an old Feistel cipher with 4 branches.
This cipher offers great {\bf flexibility} in the choice of the internal wiring
so that we can possibly make some adjustments if we do not find a property we are looking for.
Most block ciphers such as DES or AES also have this sort of flexibility in the choice of P-boxes,
arbitrary invertible matrices inside the S-box, inside the mixing layers, etc.
However there is only one such cipher (to the best of our knowledge) in which
a high degree of flexibility in the choice of internal wiring
is officially allowed, and was used
to encrypt real-life communications.
In original T-310 cipher machines
the wiring of the cipher would be changed once per year \cite{MasterPaperT310}.
Arbitrary changes in this wiring including some extremely weak ones \cite{LCKT1ucry18}
are possible, and will be entirely compatible with original historical hardware.
%

\vskip-2pt
\vskip-2pt
\subsection{On Feistel Ciphers}
\label{subsec2:Feistel}
\vskip-2pt

The question of non-linear invariants in Feistel ciphers in particular
was first studied by Knudsen and Robshaw at Eurocrypt'96 cf. \cite{GenLinear2}.  
The authors 
have actually claimed that this approach cannot or will not work for Feistel ciphers.
This type of impossibility claim is, well simply fake news,
or an incorrect interpretation of \cite{GenLinear2}.
There is no doubt non-linear invariant attacks CAN be made to work for Feistel ciphers.

One type of attack was shown in a later paper presented at Crypto 2004.
The concepts of Bi-Linear and Multi-Linear cryptanalysis
were subsequently introduced \cite{BLC,invglc,SlidesDESteach}
in order to work with Feistel ciphers with two and several branches specifically.
There are however indeed some serious difficulties in making such attacks work.
One of them is that great majority of attacks work only for a fraction of key space,
which was already clearly visible in the work of \cite{BLC}.  
In this paper and other recent work however we show that this difficulty
can be overcome entirely for T-310 cipher: the attack will work for any key (!).
For other ciphers such as DES, the problem remains, see for example Thm. 11.6 in
\cite{BackdAnn} and Section 1.1 in \cite{Zorro} and Section 5.2. in \cite{Beyne18}
which seems to suggest that one cannot have an attack which would not make some assumption on the key.
We believe indeed that for some ciphers we will maybe never find a really good attack in this respect
however our results on T-310 in this paper and other in \cite{BackdTut} suggests that one can be
more optimistic: indeed different attack for different keys can overlap and the same polynomial invariant
can be obtained in many different ways [factorization for Boolean polynomials is not unique cf. \cite{LackUnique,AidanDeg7Ucry}].

\newpage

\vskip-2pt
\vskip-2pt
\section{Specification and Boolean Polynomial Representation of T-310}
\label{subsec2:spec}

\begin{figure}[ht!]
\vskip-7pt
\begin{center}
\includegraphics*[width=5.1in,height=3.4in]{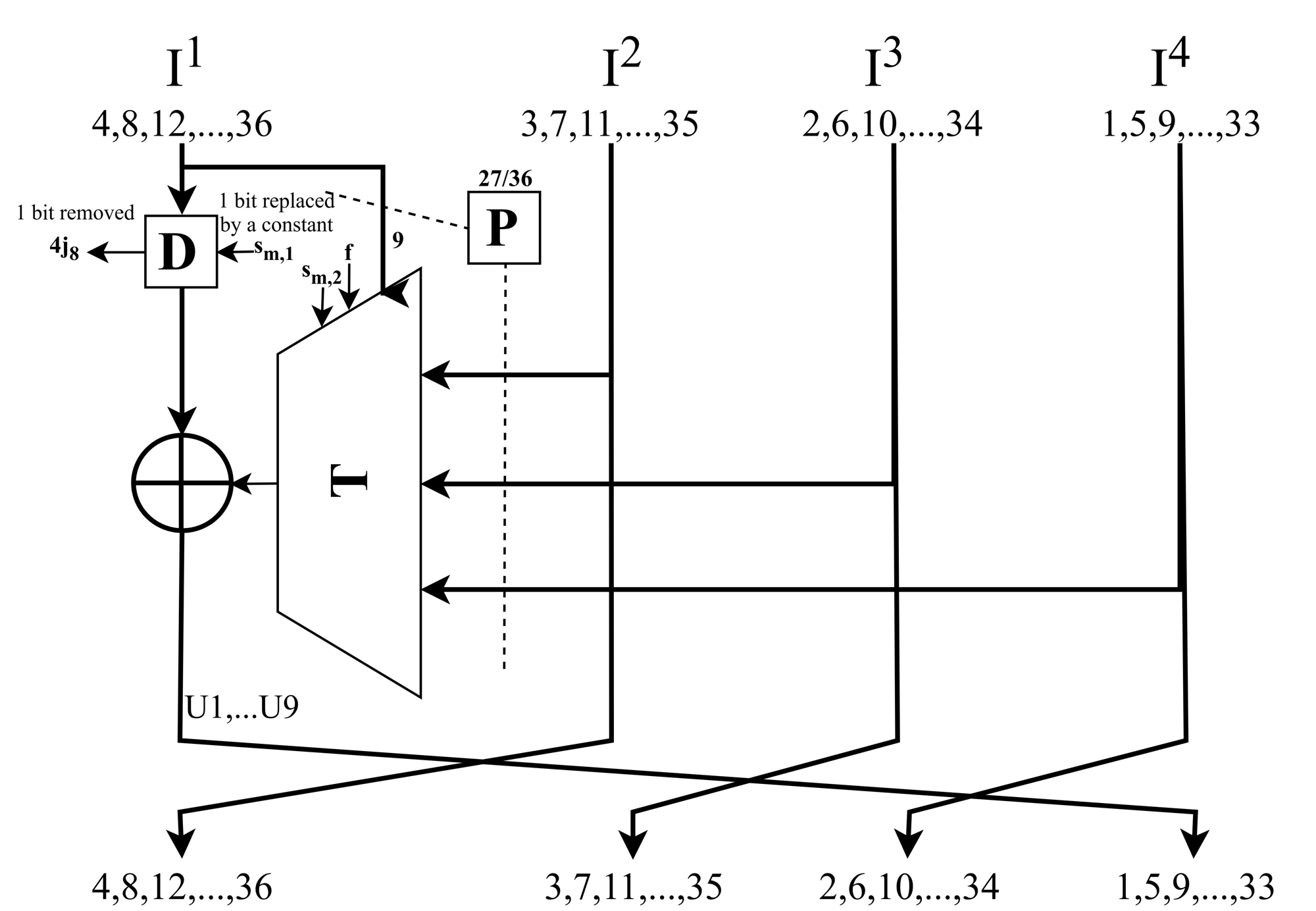}
\end{center}
\vskip-3pt
\vskip-3pt
\caption{T-310: a peculiar sort of Compressing Unbalanced Feistel scheme cf. \cite{UnbContractPata,MasterPaperT310}.
}
\label{FigContracting310KT1}
\vskip-7pt
\vskip-7pt
\end{figure}

%
The block size is 36 bits and the key has 240 bits.
%
We number the cipher state bits from 1 to 36.
The bits $1,5,9\ldots 33$ are those freshly created
in one round, cf. Fig \ref{FigContracting310KT1}.
All the input bits with numbers which are NOT multiples of 4 are shifted by 1 position, i.e.
bit 1 becomes 2 in the next round, and bit 35 becomes 36.

\vskip-5pt
\vskip-5pt
\vskip-2pt
\vskip-2pt

We denote round inputs by
$x_{1},\ldots , x_{36}$ and the
outputs will be
$y_{1},\ldots , y_{36}$.
One round of encryption can be described as 36 Boolean polynomials
of degree 6, cf. Fig. \ref{SpecFormulas} below.
Each round involves two key bits $K,L$
and one round constant bit $F$ which is derived from an IV of 61 bits
which is transmitted in the cleartext.
The initial key is defined as $s_{1\ldots 120,1\ldots 2}\in \{0,1\}^{120\cdot 2}$ which is 240 bits.
The non-linear part of our cipher are $W(),X(),Y(),Z()$,
which are four identical Boolean functions which operate on four disjoint sets of 6 variables.

%
%

\begin{figure}[ht!]
\vskip-8pt
\hskip-10pt
\hskip-10pt
\begin{center}
\hskip-10pt
\hskip-10pt
\vskip-6pt
\vskip-6pt
\includegraphics*[width=5.1in,height=1.4in]{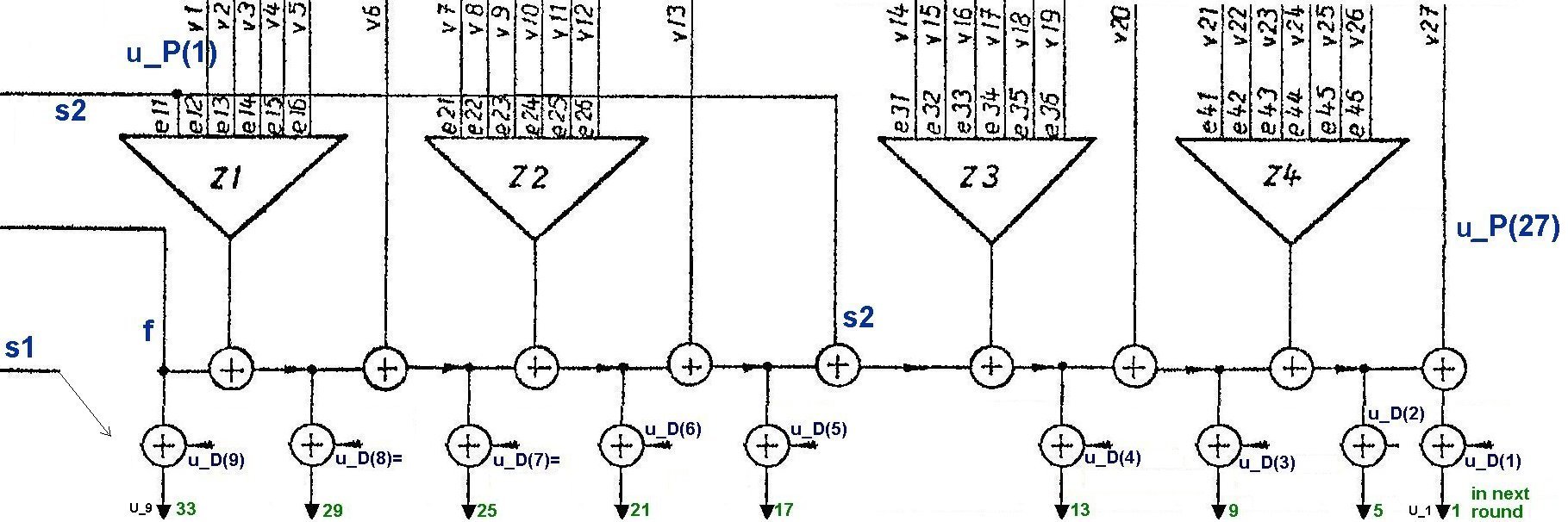}
\end{center}
\vskip-4pt
\vskip-4pt
\caption{The internal structure inside one round of T-310}
\label{FigComplicationUnit6basicFirstLook}
\end{figure}

In T-310 the specification of the internal wiring of the cipher
is called an LZS or {\em Langzeitschl\"{u}ssel}
which means a long-term key.
%
We simply need to specify
two functions
$D: \{1\ldots 9\} \to \{0\ldots 36\}$, $P:\{1\ldots 27\}\to \{1\ldots 36\}$.
For example $D(5)=36$ will mean that input bit 36 is connected
to the wire which becomes $y_{17}$ after XOR. 
Then $P(1)=25$ will mean that input 25 is connected as the 2nd input of $Z()$ also known as $v1$
in Fig. \ref{FigComplicationUnit6basicFirstLook}.
The equation (f1) means that we apply a special convention,
where the key bit $K$ is used instead of one of the $D(i)$ by specifying that $D(i)=0$.

\vskip-9pt
\vskip-9pt
\begin{align*}
\setcounter{equation}{-1}
y_{i+1}&=x_{i} \mbox{~for any~} i\ne 4k & (\mbox{~with~} 1\leq i\leq 36)~~~~~~~~~ & & \mathrm{(r0)}\\
y_{33}  &= F + x_{D(9)}  & & & \mathrm{(r1)}\\
& Z_{1} \hspace{-.28em}\stackrel{def}{=}\hspace{-.28em}Z(L,x_{P(1)},\hspace{-.20em}\ldots,\hspace{-.20em}x_{P(5)})\hspace{-.55em}\hspace{-.70em} & & & \mathrm{(z1)}\\
y_{29}  &=  F  + Z_{1} + x_{D(8)} &  & & \mathrm{(r2)}\\
y_{25}  &=  F  + Z_{1} + x_{P(6)}+x_{D(7)} & ~~~~~~~~~~~~~~~~~~~~~~~~~~~ & & \mathrm{(r3)}\\
& Z_{2} \hspace{-.28em}\stackrel{def}{=}Y(x_{P(7)},\ldots, x_{P(12)}) & & & \mathrm{(z2)}\\
y_{21}  &=  F  + Z_{1} + x_{P(6)}+Z_{2}+ &  x_{D(6)} ~~~~~~~~~~~~~~~~~~~~~~~~~&  & \mathrm{(r4)}\\
y_{17}  &=  F  + Z_{1} + x_{P(6)}+Z_{2}+ &  x_{P(13)} + x_{D(5)} ~~~~~~~~~~~~~&  & \mathrm{(r5)}\\
& Z_{3} \hspace{-.28em}\stackrel{def}{=}X(x_{P(14)},\ldots, x_{P(19)}) & & & \mathrm{(z3)}\\
y_{13}  &=  F  + Z_{1} + x_{P(6)}+Z_{2}+ & x_{P(13)} + L \hspace{-.15em}+\hspace{-.15em} Z_{3} + x_{D(4)} ~&  & \mathrm{(r6)}\\
y_{9}  &=  F  + Z_{1} + x_{P(6)}+Z_{2}+ & x_{P(13)} + L\hspace{-.15em}+\hspace{-.15em}Z_{3} + x_{P(20)} &+x_{D(3)}~~~~~~~~~~ & \mathrm{(r7)} \\
& Z_{4} \hspace{-.28em}\stackrel{def}{=}W(x_{P(21)},\ldots, x_{P(26)}) & & & \mathrm{(z4)}\\
y_{5}  &=  F  + Z_{1} + x_{P(6)}+Z_{2}+ & x_{P(13)} + L\hspace{-.15em}+\hspace{-.15em}Z_{3} + x_{P(20)} &\hspace{-.15em}+\hspace{-.15em}Z_{4} \hspace{-.15em}+\hspace{-.15em} x_{D(2)}\hspace{-.15em}   & \mathrm{(r8)}\\
y_{1}  &=  F  + Z_{1} + x_{P(6)}+Z_{2}+ & x_{P(13)} + L\hspace{-.15em}+\hspace{-.15em}Z_{3} + x_{P(20)} &\hspace{-.15em}+\hspace{-.15em}Z_{4} \hspace{-.15em}+\hspace{-.15em} x_{P(27)} \hspace{-.15em}+\hspace{-.15em} x_{D(1)}\hspace{-.15em}  & \mathrm{(r9)}\\
 & x_{0} \stackrel{def}{=}K & &  & \mathrm{(s1)}\\
F &\in \{0,1\} \mbox{~is a round constant} & \mbox{depending on a (public)} & \mbox{~IV} & (f1)\\
K &= s_{m \mbox{\scriptsize~mod~} 120,~1} & \mbox{(in encryption round ~} & m = 0,1,2,\ldots) & (k1)\\
L &= s_{m \mbox{\scriptsize~mod~} 120,~2} & \mbox{(in encryption round ~} & m = 0,1,2,\ldots) & (k2)\\
\end{align*}

\vskip-3pt
\vskip-3pt
\begin{figure}[h!]
\vskip-8pt
\vskip-8pt
\begin{center}
\vskip-8pt
\vskip-8pt
\end{center}
\vskip-8pt
\vskip-8pt
\caption{The specification of one round of T-310,
cf. also Fig. \ref{FigComplicationUnit6basicFirstLook}
and Fig \ref{FigComplicationUnit6basic}.
}
\label{SpecFormulas}
\vskip-8pt
\end{figure}

\vskip-9pt
\vskip-9pt
\subsection{Variable Naming and Polynomial Invariants}
\label{subsec2:InvNaming36}
\vskip-3pt

\begin{figure}[h!]
\vskip-3pt
\hskip-10pt
\hskip-10pt
\begin{center}
\hskip-10pt
\hskip-10pt
\vskip-3pt
\vskip-3pt
\vskip-3pt
\includegraphics*[width=5.0in,height=0.36in]{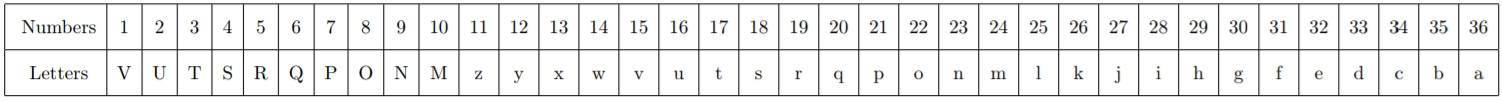}
\end{center}
\vskip-2pt
\vskip-2pt
\caption{Variable naming conventions}
\label{convert}
\end{figure}

In order for our polynomials to be short and compact we sometimes replace the 36 bits $x_{1}-x_{36}$
by single letters, cf. Fig. \ref{convert}. We avoid certain letters like $F$ used elsewhere.  
We study polynomial invariants for one round with 36 variables.
Therefore variables $x_i$ and $y_i$ are treated ``alike''
and can be called be the SAME letter, for example $x_{36}=a$ and then $y_{36}=a$ also.
If we want to avoid ambiguity, we will distinguish between the variable $a$ at input denoted by $a^i$ or just $a$,
and the same variable  at output denoted by $a^o$ or $a^\phi$ where $\phi$ is a short notation for one round of encryption.

\vskip-3pt
\vskip-3pt
\vskip-2pt
\vskip-2pt


%

\vskip-9pt
\vskip-9pt
\subsection{The Substitutions}
\label{subsec2:subs}
\vskip-3pt

When later in this paper we use some concrete LZS
(for example LZS 551 which is specified later)
this leads to replacing variables and to some simplifications,
as shown in the example below.
For example $y_{25}$ is $l$ and if $D(7)=20$ then $x_{D(7)}$ becomes $q$, etc,
and we get $l \leftarrow  F + Z1 + O + q$.
This can be interpreted as one round of encryption is equivalent
to replacing $l$ and all the other letters by our exact formulas,
for example with LZS 551 we get the following substitutions:


\vskip-9pt
\vskip-9pt
\begin{align*}
a  &\leftarrow  b & \\
b  &\leftarrow  c & \\
c  &\leftarrow  d & \\
& [\ldots] & \\
&Z1 \leftarrow Z(L,t,S,d,y,M)\\
l  &\leftarrow  F + Z1 + O + q & \\
& [\ldots] & \\
& Z4 \leftarrow W(w,u,a,h,e,n)\\
& [\ldots] & \\
V &\leftarrow F + Z1 + O + Z2 + q + L + Z3 + i + Z4 + k + j&\\
\end{align*}
\vskip-7pt
\vskip-7pt
\label{SubsEqsShorter}
\vskip-7pt


\vskip-2pt
\vskip-2pt
\section{The Fundamental Equation (FE)}
\label{sec:3}
\label{sec:FE}
\vskip-2pt

In order to break our cipher we need to find a polynomial
expression ${\cal P}$ say

\vskip-2pt
\vskip-2pt
$$
{\cal P}(a,b,c,d,e,f,g,h,\ldots) =
abcdijkl+efg+efh+egh+fgh
$$
\vskip-1pt

using any number between 1 and 36 variables
such that if we substitute in ${\cal P}$  all the variables by the substitutions defined
we would get exactly the same polynomial expression ${\cal P}$, i.e.
$
{\cal P}(\mbox{Inputs}) =
{\cal P}(\mbox{Output ANF})
$ are equal as multivariate polynomials.
We obtain: 

\begin{defi}[Compact Uni/Quadri-variate FE]
\label{defiCompactFE}
Our ``Fundamental Equation (FE)'' to solve is to make sure that sum of two polynomials like:

\vskip-3pt
\vskip-3pt
$$
FE = {\cal P}(\mbox{Inputs}) + {\cal P}(\mbox{Output ANF})
$$
\vskip-2pt

reduces to 0, or more precisely we are aiming at $\mathrm{FE}=0$ for any input,
or in other words we want to achieve a formal equality of two Boolean polynomials like

\vskip-3pt
\vskip-3pt
$$
{\cal P}(\mbox{Inputs}) +
{\cal P}(\mbox{Transformed Outputs}) = 0
$$
\vskip-3pt

or even more precisely

\vskip-2pt
\vskip-2pt
$$
{\cal P}(a,b,c,d,e,f,g,h,\ldots) =
{\cal P}(b,c,d,F+m,f,g,h,F+Z1+e,\ldots)
$$
\vskip-3pt

\end{defi}
\vskip-3pt
\vskip-3pt

where $Z1-Z4$ will be later replaced by Boolean functions $Z(),Y(),X(),W()$. 

{\bf Alternative Notation.}
There is also another notation which is more like notations used in classical invariant theory.
Instead of writing

\vskip-3pt
\vskip-3pt
$$
\mathrm{FE}=
{\cal P}(\mbox{Inputs}) +
{\cal P}(\mbox{Transformed Outputs})
$$
\vskip-3pt

we can also write:

\vskip-3pt
\vskip-3pt
$$
FE = {\cal P} + {\cal P}^{\phi}
$$
\vskip-2pt

where

\vskip-3pt
\vskip-3pt
$$
{\cal P}^{\phi} \stackrel{def}{=} {\cal P}(\phi(\mbox{Inputs})) = {\cal P}(\mbox{Transformed Outputs})
$$
\vskip-2pt

which is the same as above, and we can also write:

\vskip-3pt
\vskip-3pt
$$
{\cal P}^{i}\stackrel{def}{=}{\cal P}
$$
\vskip-2pt

\vskip-3pt
\vskip-3pt
$$
{\cal P}^{o}\stackrel{def}{=}{\cal P}(inputs^{\phi})
$$
\vskip-2pt

where $\phi$ is the transformation induced by 1 round of encryption and
where $\phi(\mbox{Inputs})$ denotes a sequence of 36 polynomial expressions
of output-side variables $(a,\ldots V)$ 
expressed as Boolean function of the 36 input-side variables
[with some extra variables such as secret key variables].
For example the variable $a$ is replaced by polynomial $b$ and $d$ by $F+m$.
In other words they are written as formal polynomials in $B_{36}$
corresponding to the ANF expressions of one round of encryption
(and as a function of inputs of this round).
Our usage of exponents is similar as in the mathematical (Hilbertian) invariant theory.
Our exponents can be simply interpreted as transformations on polynomials,
or more precisely as operations belonging to a certain group of transformations
acting on a set of Boolean polynomials ${\cal P}$ or $A$ or other say $(azM+b) \in B_{36}$ where $B_{36}$ is the
precise ring of all Boolean polynomials in 36 variables named $a-z$ and $M-V$ as in this paper.
The notation ${\cal P}^{\phi}$ is very elegant and unhappily {\bf ambiguous} or not always well-defined.
This is because in general $\phi$ depends also on $F$ and various key bits.
Then it happens that ${\cal P}^{\phi}$ is likely to be unique nevertheless:
we are aiming at computing ${\cal P}^{\phi}$ primarily and precisely in cases where the result,
the transformed and substituted polynomial ${\cal P}^{\phi}$ is such
that the final result ${\cal P}^{\phi}$ does NOT depend on the variables $F,K,L$ (!).
This may and will become possible when our polynomial ${\cal P}$ is particularly well chosen.

In the next step, $Z$ will be represented by an Algebraic Normal Form (ANF)
with 64 binary variables which are the coefficients of the ANF of $Z$,
and there will be several equations,
and four {\bf instances} $Z,Y,X,W$ of the same $Z$:

\begin{defi}[A Multivariate FE]
\label{defiRewriteFEZ00}
At this step 
we will rewrite FE as follows.
We will replace Z1 by:

\vskip-3pt
\vskip-3pt
$$
Z \leftarrow Z00+Z01*L+Z02*j+Z03*Lj+\ldots +Z62*jhfpd+Z63*Ljhfpd
$$
\vskip-1pt

Likewise we will also replace $Z2$:
\vskip-3pt
\vskip-3pt
$$
Y \leftarrow Z00+Z01*k+Z02*l+Z03*kl+\ldots +Z62*loent+Z63*kloent
$$
\vskip-1pt
\noindent
and likewise for $X=Z3$ and $W=Z4$ and the coefficients $Z00\ldots Z63$
will be the same inside $Z1-Z4$, however the subsets of 6 variables
chosen out of 36 will be different in $Z1-Z4$.
Moreover, some coefficients of ${\cal P}$ may also be variable.
\end{defi}
\vskip-3pt


%
\noindent
In all cases, all we need to do is to solve our $\mathrm{FE}$ equation for $Z$,
i.e. determine 64 binary 
variables $Z00\ldots Z63$.
This formal algebraic approach, if it has a solution,
still called $Z$
for simplicity, 
or $({\cal P},Z)$
will {\bf guarantee} that our invariant ${\cal P}$ holds for 1 round.
This is, and in this paper we are quite lucky,
IF this equation does not depend on three bits $F,K,L$.
This is the discovery process of \cite{BackdTut}.

In this paper we do NOT use this process.
We will work  by {\bf attack hopping}.
From one attack we will derive the existence of another attack
on a different cipher wiring (i.e. a different LZS).
Thus we completely avoid all the most difficult questions in \cite{BackdTut}:
Do such equations have any solutions?
If they have, can the solution be the same for several permutations simultaneously
so that our attack can work in presence of key bits, round constants, IV bits etc?
We concentrate on transposing some working attacks
to another cipher setup or where the Boolean function is modified.
No systematic method to study all possible invariant attacks is known and possibly such method does not exist
due to double-exponential explosion in the number of possible polynomial invariants.

\newpage

\vskip-2pt
\vskip-2pt
\section{KT1 Keys and Higher Degree Invariant Attacks}
\label{sec:KT1Quadratic}
\label{sec:4}
\label{SimpleInvP20Cycle9BiasedFEHomQuadBakdPaper551DifferentOpticHigherDegree4}
\vskip-2pt

Many polynomial invariants on block ciphers published so far are of low degree,
mainly of degree 2 \cite{BLC,BackdTut,TodoNL18}
and finding any such invariants was quite difficult.
Many examples presented are also quite artificial because they have little to
do with any encryption systems used to encrypt real-life communications.
The East German government cryptologist have
mandated that for an LZS to be approved for ``official'' use,
it must satisfy a certain very complex specification called KT1 which
takes one full page to describe, cf. Appendix B in \cite{MasterPaperT310}.
Therefore we expect that it is substantially harder to find an attack on a KT1 key.
Our starting point will be some simple invariants from \cite{BackdTut}
which actually work for a genuine KT1 key,
something considered very hard to do until recently.
There are approximately $2^{83}$ possible KT1 keys, cf. Section 5.4. of \cite{LCKT1ucry18}
while there are approximately $2^{170}$ possible\footnote{
A quick estimation is ${36\choose 27}\cdot 27! \cdot {36\choose 9}\cdot 9!\cdot 36\approx 2^{170}$.}
LZS keys.
The reader can therefore imagine that the attacks we present here for KT1 keys are substantially harder to find
than most of earlier attacks on T-310 such as found in \cite{BackdTut}:
numerous technical constraints must be satisfied simultaneously.

It is known that about 3 $\%$ of KT1 keys can are extremely weak
w.r.t. Linear Cryptanalysis, cf. \cite{SlideLCWeakCanAttack310}.
Inside the remaining 97 $\%$ of cases, can a KT1 key be vulnerable to a non-linear invariant attack?
The answer is that in some cases yes, as we are going to show later.

\vskip-3pt
\vskip-3pt
\subsection{Starting Point: A Peculiar Quadratic Attack on a KT1 Key}
\label{subsec4:linear}

\vskip-2pt
\vskip-2pt
\begin{verbatim}
551: P=17,4,33,12,10,8,5,11,9,30,22,24,20,2,21,34,1,25,
13,28,14,16,36,29,32,23,27 D=0,12,4,36,16,32,20,8,24
\end{verbatim}
\vskip-3pt

\vskip-2pt
\vskip-2pt
\vskip-2pt
$$
{\cal P}=eg+fh+eo+fp+gm+hn+mo+np
$$
\vskip-2pt

\noindent
with this short ${\cal P}$ the Fundamental Equation FE will have very few terms:

\vskip-2pt
\vskip-2pt
$$
{\cal P}(a,b,c,d,e,f,g,h,\ldots,V) =
{\cal P}(b,c,d,F+m,f,g,h,F+Z+O,$$
\vskip-2pt
\vskip-2pt
$$
\ldots,F+Z+O+Y+q+L+X+i+W+j+K)
$$
\vskip-3pt

\vskip-2pt
\vskip-2pt
$$
(Y+m)(g+o)=0
$$
\vskip-1pt

\noindent
and one solution which makes our cipher weak is $Z=1+d+e+f+de+cde+def$.
The fact that our $\mathrm{FE}$ contains none of $F,K,L$ implies that our polynomial ${\cal P}$ is an invariant
which works for any key and any $IV$
and for any number of rounds.

\begin{figure}[h!]
\vskip-2pt
\vskip-2pt
\hskip-10pt
\hskip-10pt
\begin{center}
\hskip-10pt
\hskip-10pt
\vskip-3pt
\vskip-3pt
\includegraphics*[width=5.1in,height=1.6in]{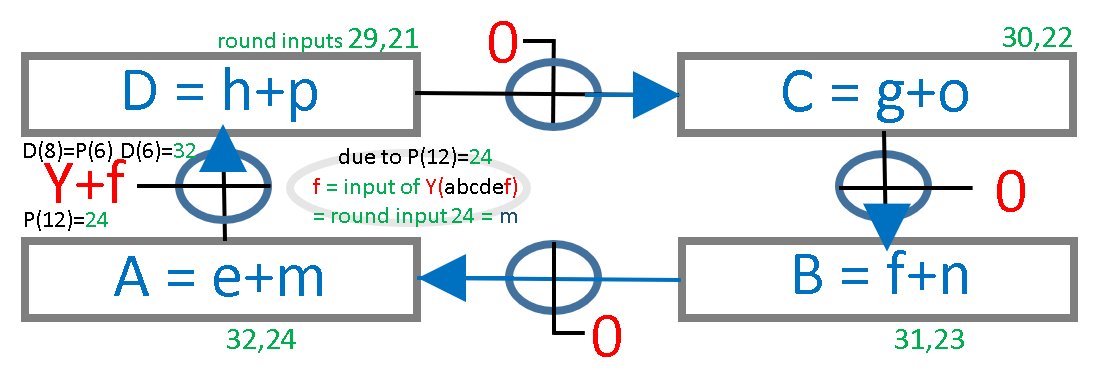}
\end{center}
\vskip-1pt
\vskip-1pt
\caption{Connections between that four linear factors used in our invariant attack ${\cal P}=ABCD$.}
\label{CycleFigABCDFor551}
\vskip-3pt
\vskip-3pt
\end{figure}

It may seems that $(Y+m)(g+o)=0$ was obtained by some sort of accident,
later we will see why this attack works,
cf. later Thm. \ref{ThmKT1Cycle551Like2}
and Fig. \ref{CycleFigABCDFor551}.
Before that we need to make a number of observations.
We abstract our $\mathrm{FE}$ from notations used for the whole cipher,
and look at the inputs of our Boolean function.
Then our $\mathrm{FE}$ can be re-written as the following ``annihilation requirement'' using $Z+f$,
where $f$ is an input of $Z$, and $Z+f$ must be annihilated by a linear term $d+e$,
and $d,e$ are also inputs of $Z$:

\vskip-4pt
\vskip-4pt
$$
(Z+f)(d+e)=0
$$

\vskip-3pt
\vskip-3pt
\subsection{An Essential Insight}
\label{subsec4:decomposeACBD}
It is easy to see that ${\cal P}$ is irreducible and
no linear attack exists for this cipher setup.
Interestingly we have ${\cal P}=AC+BD$ where:


\vskip-2pt
\vskip-2pt
$$
\begin{cases}
A\stackrel{def}{=}  (e + m) = x_{32}+x_{24}
\cr
B\stackrel{def}{=}  (f + n) = x_{31}+x_{23}
\cr
C\stackrel{def}{=}  (g + o) = x_{30}+x_{22}
\cr
D\stackrel{def}{=}  (h + p) = x_{29}+x_{21}
\end{cases}
$$
\vskip-1pt

\noindent
Now do $A,B,C,D$ have any concrete significance for our cipher?
To see this let us consider a yet simpler case when $Z(a,b,c,d,e,f)=f$.
Then it is possible to check that our cipher would have 4R linear invariant $D\to C\to B\to A\to D$
which however is totally absent when $Z=1+d+e+f+de+cde+def$.

If we go back to a non-linear Boolean functions,
inside $D\to C\to B\to A\to D$ we have 3 trivial transitions which are always true
and one ``impossible'' transition $A\to D$ which is no longer correct
(unless $Y$ is equal to its last input).

\vskip-3pt
\vskip-3pt
\subsubsection{Eliminating the Output of Boolean Function $Y$}
\label{subsec4:proofTransitionAtoD}

Interestingly, on Fig. \ref{CycleFigABCDFor551} we ``almost'' have $A\to D$ transition,
the difference $A^i+D^o$ is actually equal to the output of Boolean function
Z2 which is also denoted as $Y$ plus an affine function.
This is due to internal wiring in LZS 551 and can be checked either by
applying equations of one round in Section \ref{subsec2:spec},
or much more easily by simply summing bits on a path going from in output 21 to 29 through
bits numbered $D(6),D(8),P(6)$ and the output $Y$ precisely in Fig. \ref{FigComplicationUnit6basic}.
Here bits $D(8),P(6)$ cancel, $D(6)$ connects to input bit 32, and bit 32 goes to 6-th input $f$ of $Y()$.
Now in some sense in our attacks we are allowed to consider that $Y+f$ behaves as it was zero.
More precisely, we are going to see that $Y+f$ and later $Y+e$
can be annihilated in different ways, more or less trivial
in different attacks inside this paper.
We will first revisit our attack ${\cal P}=AC+BD$ and show why and when it works.

Looking at Fig. \ref{CycleFigABCDFor551} we are now able to abstract the attack on key 551
which was first discovered in \cite{BackdTut} and we can propose a more general result and attack.
Our attack corresponds to a sum of products of all polynomials lying on two disjoint sub-cycles of length 2
which form a partitioning of the whole cycle.

\vskip-3pt
\vskip-3pt
\subsubsection{A More General Result Beyond LZS 551}
\label{subsec4:proofcycle4for551}

An important insight is that in order for $AC+BD$ to be invariant we need to show that:

$$
{\cal P}^{o} = A^o C^o +B^o D^o = A^i C^i +B^i D^i = {\cal P}^{i}
$$

and due\footnote{Which in turn is due to bit shifts $k\to k+1$ for any $k\ne 4j$ inside our cipher,
cf. Fig. \ref{FigContracting310KT1}.}
to the structure of our cycle we always have one transition which is trivial:

$$
B^i D^i = A^o C^o
$$

therefore it remains to show that for any input for 1 round we have:

$$
B^o D^o = A^i C^i
$$

moreover we have another trivial property $C^i = B^o$ and our equation boils down to:

$$
C^i (D^o + A^i) = 0
$$

This is and will be achieved in our [previous and generalized] attack
using two neatly distinct assumptions.
First we are going to make sure through the cipher wiring that
we have the transition with addition of $Y+f$ or similar as depicted
on the left hand side of Fig. \ref{CycleFigABCDFor551}, we aim at having:

$$
(D^o + A^i) = Y(\mbox{its 6 inputs}) + \mbox{(some linear combination of these 6 inputs)}
$$

Then we are going to make sure that this affine modification of our Boolean function $Y+f$ or similar
will be annihilated by polynomial $C^i$.
More precisely we show that: 

\newpage
\vskip-3pt
\vskip-3pt
\setcounter{theorem}{2}
\begin{theorem}[KT1 compatible invariant]
\label{ThmKT1Cycle551Like2}
For each long term key s.t. $D(8)=P(6)$,
$D(6)=32$ and $P(10)=30$, $P(11)=22$ and $P(12)=24$,
if the Boolean function 
is such that $(Z+f)(d+e)=0$
and for any short term key on 240 bits, and for any initial state on 36 bits,
we have the non-linear invariant
$\mathcal{P}=AC+BD$ true with probability 1.0 for any number of rounds.
\end{theorem}
\vskip-2pt

\begin{figure}[ht!]
\vskip-2pt
\vskip-2pt
\hskip-10pt
\hskip-10pt
\begin{center}
\hskip-10pt
\hskip-10pt
\vskip-3pt
\vskip-3pt
\includegraphics*[width=5.1in,height=2.6in]{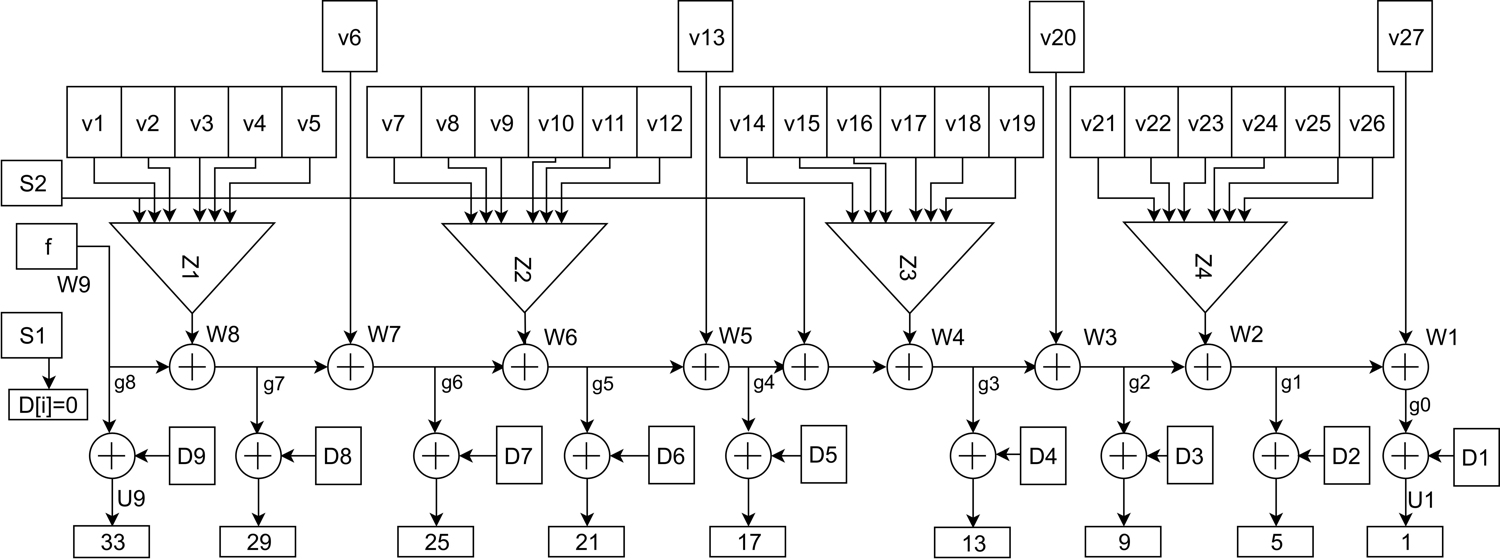}
\end{center}
\vskip-3pt
\vskip-3pt
\caption{The internal structure of one round of T-310 block cipher.
}
\label{FigComplicationUnit6basic}
\vskip-3pt
\vskip-3pt
\end{figure}

\noindent\emph{Proof:}
We first show how $(D^o + A^i)$ can be equal to a sum of $Y()$ and some of its inputs.
We XOR two equations (r2) and (r4) from the formulas of Fig. \ref{SpecFormulas}
in Section \ref{subsec2:spec} which specify our cipher:

$$
D^o = y_{29}+y_{21}  =
F  + Z1 + x_{D(8)} + F  + Z1 + x_{P(6)}+Z2+ x_{D(6)}
$$

then we have $D(8)=P(6)$ and two terms cancel and we get:

$$
D^o =
x_{D(8)} + x_{P(6)}+ Z2 + x_{D(6)} = Y() + x_{D(6)}
$$

and we check that $D(6)=32$ and therefore we have:

$$
D^o =
Y() + x_{32}
$$

which knowing that $A^i=x_{24}+x_{32}$ becomes in turn

$$
D^o =
A^i + Y()+x_{24}
$$

where $x_{24}$ is the same as last input $f$ for this Boolean function, due to $P(12)=24$.
The same result can be obtained following the path going from in output 21 to 29
through bits numbered $D(6),D(8),P(6)$ and the output $Y$ in Fig. \ref{FigComplicationUnit6basic}.
Finally we also check that $(Y+x_{24})C^i=0$ is the same as $(Z+f)(d+e)=0$
due to $C^i=x_{30}+x_{22}$ and $P(10)=30$, $P(11)=22$ and $P(12)=24$.
$\qed$

\vskip-3pt
\vskip-3pt
\subsubsection{Downgrading Our Invariant:}
\label{subsec4:simplerlinear}
Now we show that our non-linear attack {\bf hides the existence} of a yet simpler attack
with a degenerated Boolean function.
To shows this we observe that input $f$ of $Y$ is $P[12]=m$,
i.e. last input of $Z2$ is precisely connected to $m$ with LZS 551.
What happens if $Y=m$ i.e. $Z=f$? We have another degenerated solution of our $\mathrm{FE}$ being $(Y+m)(g+o)=0$
and we have in fact found a root for a {\bf proper factor} $(Y-m)=0$ of our previous $\mathrm{FE}$.

\vskip1pt
\noindent
{\bf Next Step.} An interesting question is can we do this in a reverse direction?
Find a cipher setup where the $\mathrm{FE}$ is a multiple of $(Y-m)(g+o)=0$ ?
Yes! 

%

\vskip-3pt
\vskip-3pt
\subsection{Construction of Higher Degree Invariants}
\label{subsec4:degree4}
Such invariants if they exist are NOT uniquely determined,
they may depend on the choice of the Boolean function $Z$ in the previous step (!).
For the current pair $LZS,Z$ as above we found that the following ${\cal P}$ of degree 4 also works:

\vskip-2pt
\vskip-2pt
\vskip-2pt
$$
{\cal P}=
efgh+fghm+eghn+ghmn+efho+fhmo+ehno+hmno+
$$
\vskip-2pt
\vskip-2pt
$$
efgp+fgmp+egnp+gmnp+efop+fmop+enop+mnop
$$
\vskip-1pt

where in fact ${\cal P}=ABCD$.
Interestingly, {\bf no other} invariants being polynomials in $A,B,C,D$ exist.
We conjecture however that no invariants other than $AC+BD$ (irreducible)
and $ABCD$ (not irreducible) exist for 1 round\footnote{
For 2 rounds we have closely related invariants $AC\to BD\to AC$
with $AC$ and $BD$ being invariants for 4 rounds. All these do not use $F,K,L$ either.
}
and probably not in general\footnote{In fact it is hard to be sure, no method to explore
all possible invariants with 36 variables at higher degrees is known and possibly such method does not exist}.
For sure we have verified that no linear invariants exist here for LZS 551 and $Z=1+d+e+f+de+cde+def$.

\subsubsection{Study of $\mathrm{FE}$} 
\label{subsec4:FE}
If ${\cal P}=ABCD$ what is the $\mathrm{FE}$?
A quick computation gives
%

\vskip-2pt
\vskip-2pt
$$
mBCD =YBCD
$$
\vskip-2pt

This decomposition implies that any solution $Z$ which is a solution to the previous $\mathrm{FE}$ will also work here but
{\bf NOT vice versa}.


\vskip-0pt
\vskip-0pt
\subsection{A More Autonomous Example of An Invariant of Degree 4}
\label{subsec4:betterdeg4}
\label{SimpleInvP20Cycle9BiasedFEHomQuadBakdPaper551DifferentOpticHigherDegree4Better}
\vskip-2pt

Until now we have seen that a weak cipher with linear invariants $A,B,C,D$
shared the same non-linear invariants with a cipher
where the only attacks are the non-linear ones $AC+BD$ and $ABCD$.
Is it possible to {\bf remove} the first attack and keep the second?
Yes and it is requires minimal change.
We recall that $AC+BD$ will be an invariant
each time our Boolean function satisfies:

\vskip-4pt
\vskip-4pt
$$
YC=mC
$$
\vskip-1pt

and in order for $ABCD$ to be an invariant, $Z$ needed to satisfy:

\vskip-4pt
\vskip-4pt
$$
mDCB =YDCB
$$
\vskip-2pt

All we have to do now is to find a solution $Y$ which satisfies
one $\mathrm{FE}$ and not the other $\mathrm{FE}$!
For example we can find a solution to an alternative equation:

\vskip-4pt
\vskip-4pt
$$
mB =YB
$$
\vskip-2pt

which is different than the most trivial solution $Y(......)= m$ and which
then will satisfy only the first equation $mDCB =YDCB$. 
One example of such solution is $Y=1+n+nm+f+mf$ using the same variable names.
A better solution is to modify LZS 551 very slightly:
we just need to make sure that letters $f$ and $n$ are actually inputs of $Y$.
Only two modifications are needed. Here is a solution found by a SAT solver:

\vskip-2pt
\vskip-2pt
\begin{verbatim}
558: P=17,4,33,12,10,8,23,24,31,25,16,10,20,2,21,34,
1,25,13,28,14,16,36,29,32,23,27 D=0,12,4,36,16,32,20,8,24
Z(a,b,c,d,e,f)=1+a+ab+c+bc
\end{verbatim}
\vskip-3pt

\noindent
We have checked that no other invariants at degree up to 3 exist with all the 36 state variables
for $1,2,3,4,5,\ldots$ and various numbers of rounds. 
All simple invariants were removed with the new Boolean function and only $ABCD$ is left.


\vskip-6pt
\vskip-6pt
\subsection{A Yet Stronger Example}
\label{subsec4:betterdeg4_}
\label{SimpleInvP20Cycle9BiasedFEHomQuadBakdPaper551DifferentOpticHigherDegree4Better6Vars}
\vskip-2pt

One step further, we try to find a non-trivial (proper) solution to:

\vskip-5pt
\vskip-5pt
$$
mBC =YBC
$$
\vskip-3pt

Here we will need to use as inputs of $Z$ all the 5 variables which appear in this equation.
%
%
%
%
A nice trick to quickly find a solution which is a ``proper'' root of $(m+Y)BC$
is to first create a new equation $FE'$ which implies the previous one by multiplying both sides by $B$,
yet at the same time $FE'$ actually imposes the presence of the two variables $f,n$ in $B$, and
another $6-th$ variable, for example:

\vskip-5pt
\vskip-5pt
$$
(Y+m)(f+n)=(Rnf+Rf)go
$$
\vskip-1pt


Again just one invariant $ABCD$ remains after changing the Boolean function
but now our Boolean function must use 6 quite specific variables which
must be connected to inputs of $Z2=Y$. One possible solution is as follows:

\vskip-2pt
\vskip-2pt
\begin{verbatim}
550: P=17,4,33,12,10,8,22,23,24,31,30,20,20,2,21,34,
1,25,13,28,14,16,36,29,32,23,27 D=0,12,4,36,16,32,20,8,24
Z(a,b,c,d,e,f)=1+b+c+d+aef+abef
\end{verbatim}
\vskip-3pt




\vskip-2pt
\vskip-2pt
\section{A Transposed Attack with Two Cycles}
\label{subsec5:94}
\vskip-2pt

We want to further modify this attack in order to make it stronger.
We consider a method (found with help of our research assistant Matteo Abbondati)
to imitate or transpose the current attack by designing
two independent cycles of length 4 based on the same principle
as on Fig. \ref{CycleFigABCDFor551}.
We would like two such cycles to exist simultaneously for the same cipher wiring. We define:

\begin{figure}[h!]
\vskip-7pt
\vskip-7pt
\hskip-10pt
\hskip-10pt
\begin{center}
\hskip-10pt
\hskip-10pt
\vskip-3pt
\vskip-3pt
\includegraphics*[width=5.1in,height=1.6in]{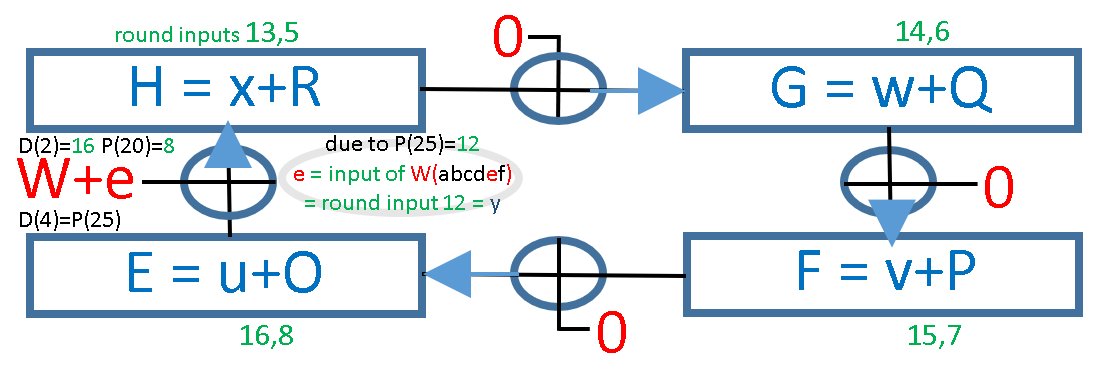}
\includegraphics*[width=5.1in,height=1.6in]{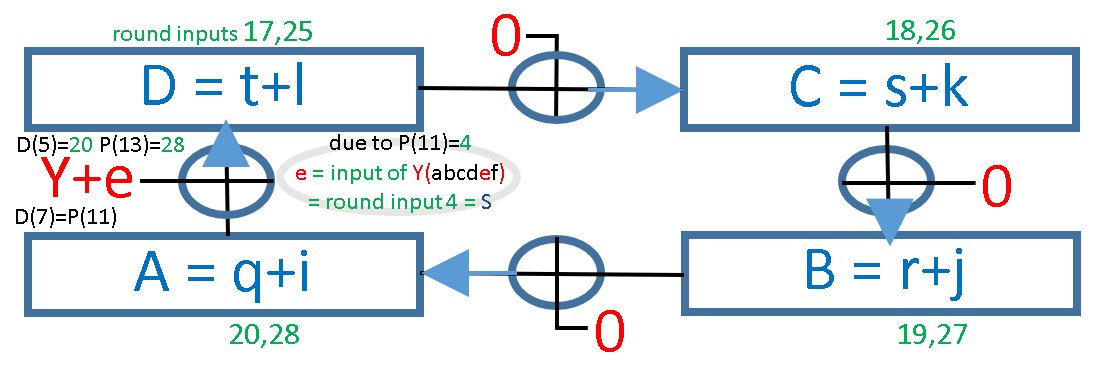}
\end{center}
\vskip-1pt
\vskip-1pt
\caption{Two cycles with 4+4 linear factors
and 2 invariant attacks with ${\cal P}=ABCD$ and
${\cal P}=EFGH$. }
\label{CycleFigABCDEFGH443}
\vskip-3pt
\vskip-3pt
\end{figure}

\vskip-2pt
\vskip-2pt
$$
\begin{cases}
A\stackrel{def}{=}  (q + i) = x_{20}+x_{28}
\cr
B\stackrel{def}{=}  (r + j) = x_{19}+x_{27}
\cr
C\stackrel{def}{=}  (s + k) = x_{18}+x_{26}
\cr
D\stackrel{def}{=}  (t + l) = x_{17}+x_{25}
\cr
E\stackrel{def}{=}  (u + O) = x_{16}+x_{8}
\cr
F\stackrel{def}{=}  (v + P) = x_{15}+x_{7}
\cr
G\stackrel{def}{=}  (w + Q) = x_{14}+x_{6}
\cr
H\stackrel{def}{=}  (x + R) = x_{13}+x_{5}
\end{cases}
$$
\vskip-1pt

\noindent
Here one example of LZS constructed to satisfy the conditions of Fig. \ref{CycleFigABCDEFGH443} is:

\vskip-2pt
\vskip-2pt
\begin{verbatim}
443: 9,19,33,7,10,3,18,26,17,30,4,25,28,2,21,34,1,11,
15,8,14,6,13,29,12,5,27 D=24,16,36,12,20,32,4,8,28
\end{verbatim}

Here the inputs of $W$ are in order
$$
{\color{red}18},
{\color{red}26},
{\color{blue}17},
{\color{black}30},
{\color{green}4},
{\color{blue}25}
$$
and inputs of $Y$ are in order
$$
{\color{red}14},
{\color{red}6},
{\color{blue}13},
{\color{black}29},
{\color{green}12},
{\color{blue}5}
$$

A quick computation shows that the Fundamental Equation
for ${\cal P}=ABCD$ is exactly equal to:

$$
B C D (S+Y()) =0
$$

which can be rewritten as, if we call $a,b,c,d,e,f$ the inputs of $Y$:
$$
(x_{19}+x_{27})(a+b)(c+f) (e+Y(a,b,c,d,e,f)) =0
$$

For ${\cal P}=EFGH$ the Fundamental Equation is equal to:

$$
F G H (y+W()) =0
$$

which can be rewritten as, if we call $a,b,c,d,e,f$ the inputs of $Y$:

$$
(x_{15}+x_{7}) (a+b)(c+f) (e+W(a,b,c,d,e,f)) =0
$$

From here it is easy to see that we can make both equations vanish simultaneously if we have:

$$
\forall_{a,b,c,d,e,f}~~
(a+b)(c+f) (e+Z(a,b,c,d,e,f)) =0
$$

for our Boolean function used twice as both $Y$ and $W$.
Here is an example of a Boolean function which works:
$Z=1+a+bc+abc+d+abd+acd+bcd+be+abe+ce+bce+abce+ade+abde+acde+f+af+acf+bcf+adf+bdf+abdf+ef+bef+abef+cef+bcef+adef+acdef$.
Moreover, a random Boolean function will satisfy this equation with probability $2^{16}$,
cf. Thm C.2 in Appendix C in \cite{BackdAnn} or Section 5 in \cite{AidanDeg7Ucry}.
In this attack, it easy to see that in addition to two invariants of degree 4,
an invariant of degree 8 also exists. More precisely,
a product of two invariants for a block cipher is always an invariant.
Therefore
$
{\cal P}=ABCDEFGH
$
is also an invariant with LZS 443 and any Boolean function which satisfied our FE equation as above.

\vskip-3pt

\vskip-2pt
\vskip-2pt
\section{Removing Lower-degree Invariants and Increasing the Degree in Our Annihilation Requirement}
\label{subsec5:95}
\vskip-2pt

We have further modified the LZS and the Boolean function so that 
invariants of degree 4 disappear
and only one invariant of degree 8 remains.

$$
{\cal P}=ABCDEFGH
$$

this is achieved for a modified cipher wiring as follows:

\vskip-2pt
\vskip-2pt
\begin{verbatim}
444: P=17,1,33,2,10,3,18,26,19,27,36,5,16,32,21,34,8,
25,13,28,14,6,15,7,12,23,30 D=24,20,4,12,8,32,36,16,28
\end{verbatim}
\vskip-3pt

where the inputs of $W$ are in order
$$
{\color{red}18},
{\color{red}26},
{\color{blue}19},
{\color{blue}27},
{\color{green}36},
{\color{magenta}5}
$$
and
inputs of $Y$ are in order
$$
{\color{red}14},
{\color{red}6},
{\color{blue}15},
{\color{blue}7},
{\color{green}12},
{\color{magenta}23}
$$

\begin{figure}[h!]
\vskip-2pt
\vskip-2pt
\hskip-10pt
\hskip-10pt
\begin{center}
\hskip-10pt
\hskip-10pt
\vskip-3pt
\vskip-3pt
\includegraphics*[width=5.1in,height=3.5in]{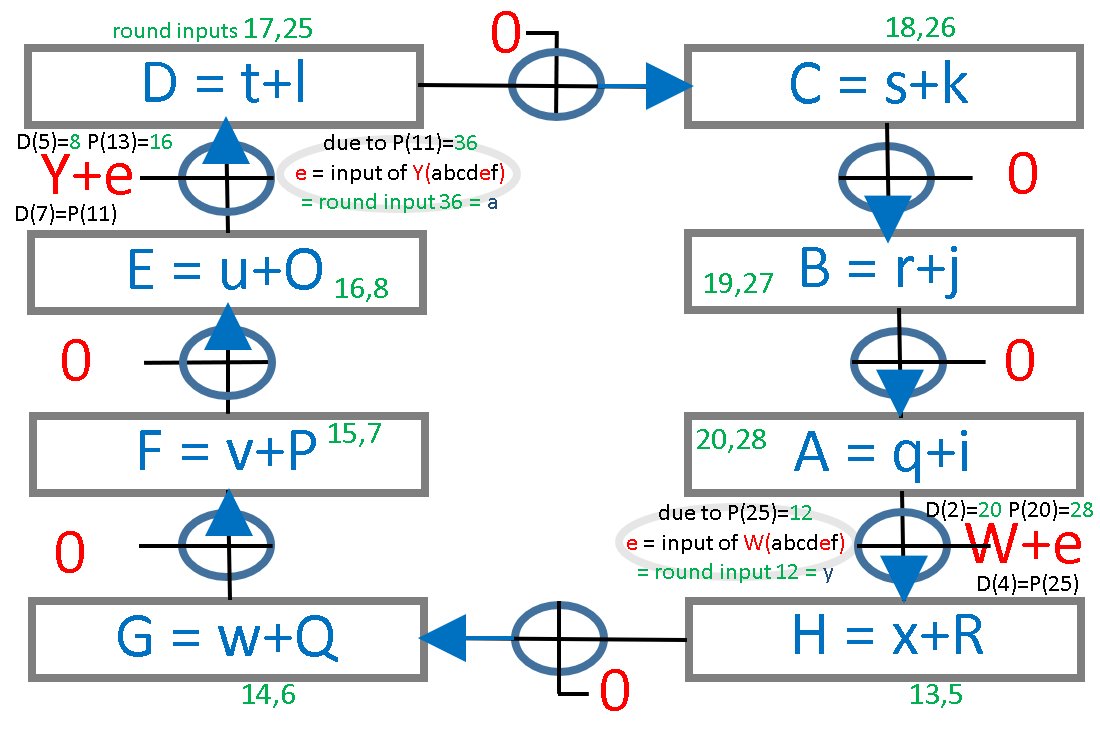}
\end{center}
\vskip-1pt
\vskip-1pt
\caption{Connections between the 8 linear factors used in our second attack with ${\cal P}=ABCDEFGH$.}
\label{CycleFigABCDEFGH444}
\vskip-3pt
\vskip-3pt
\end{figure}

The main principle is that we disconnect the two cycles in Fig. \ref{CycleFigABCDEFGH443}
and connect them to make one single cycle.
This is done by literally exchanging the two values of $D(5),P(13)$ with
$D(2),P(20)$ which has the effect to combine the two cycles,
cf. Fig. \ref{CycleFigABCDEFGH444}.
The constraints $D(7)=P(11)$ and $D(4)=P(25)$ remain unchanged.
Then we verify that the new wiring gives a bijective round function (which is not obvious).

Now we compute the Fundamental Equation (FE).

\vskip-3pt
\vskip-3pt
$$
\mathrm{FE}=
{\cal P}(\mbox{Inputs}) +
{\cal P}(\mbox{Transformed Outputs})
=
{\cal P} + {\cal P}^{\phi}
$$
\vskip-3pt

We start by observing that, following the path
from output 13 to 5 in Fig. \ref{FigComplicationUnit6basic},
or adding together the equations (r6) and (r8)
in Fig. \ref{FigComplicationUnit6basicFirstLook} we get:

$$
H^o=y_{13}+y_{5}= x_{D(4)} + x_{P(20)} + W(.) + x_{D(2)}=
(W()+x_{P(25)}) + (x_{28}+x_{20})
=
(W+e)(.)+A^i
$$
\vskip-1pt

Here by $(W+e)$ we denote a modified Boolean function with addition of the $5$-th input bit.
Then following the path
from output 25 to 17 in Fig. \ref{FigComplicationUnit6basic},
or adding together the equations (r3) and (r5)
in Fig. \ref{FigComplicationUnit6basicFirstLook} we get:
\vskip-4pt
\vskip-4pt
$$
D^o=y_{25}+y_{17}=
x_{D(7)} + Y(.) + x_{P(13)} + x_{D(5)}=
(Y(.)+x_{P(11)}) + (x_{16} + x_{8})=
(Y+e)(.)+E^i
$$
\vskip-1pt

We have then:

\vskip-3pt
\vskip-3pt
$$
\mathrm{FE}=
{\cal P} + {\cal P}^{\phi}
=
A^iB^iC^iD^iE^iF^iG^iH^i +
A^oB^oC^oD^oE^oF^oG^oH^o
=
$$
\vskip-3pt

\vskip-3pt
\vskip-3pt
$$
\mathrm{FE}=
A^iB^iC^iD^iE^iF^iG^iH^i +
B^iC^iD^i((Y+e)(.)+E^i)
F^iG^iH^i((W+e)(.)+A^i) =
$$
\vskip-3pt

Now we have only input-side variables and we can drop the $^i$ notations:

\vskip-3pt
\vskip-3pt
$$
\mathrm{FE}=
BCDFGH
\left (~~
AE ~~+~~
[(W+e)(.)+A]
[(Y+e)(.)+E]
~~
\right)
$$
\vskip-3pt

which can be rewritten as:

\vskip-3pt
\vskip-3pt
$$
\mathrm{FE}=
BCDFGH
\left (~~
E(W+e)(.)+
A(Y+e)(.)+
(W+e)(.)(Y+e)(.)
~~
\right)
$$
\vskip-3pt

We will now discard factors such as $D$ which have variables not used as inputs of our Boolean functions $W,Y$.
In contrast knowing that inputs of $W$ are
$
{\color{red}18},
{\color{red}26},
{\color{blue}19},
{\color{blue}27},
{\color{green}36},
{\color{magenta}5}
$
the polynomials $B,C$ are legitimate potential annihilating factors for $W$.
Similarly, the inputs of $Y$ are
$
{\color{red}14},
{\color{red}6},
{\color{blue}15},
{\color{blue}7},
{\color{green}12},
{\color{magenta}23}
$
and $F,G$ can be used.
It is then easy to see that one way to annihilate the $\mathrm{FE}$ equation above is
to have simultaneously:

\vskip-5pt
\vskip-5pt
$$
\begin{cases}
BC(Y+e)=0\cr
FG(W+e)=0\cr
\end{cases}
$$
\vskip-1pt

which also will insure the annihilation of our third term with
$BCFG(W+e)(Y+e)=0$.
Knowing that these Boolean function are identical
We obtain two identical annihilation
requirements being exactly:

$$
(Z+e)(a+b)(c+d)=0
$$

One solution is

$$
Z=
fedcb+fedca+fedc+fecba+fecb+feca+fec+feba+feb+fea+fe
$$
\vskip-5pt
\vskip-5pt
$$
+fdb+fd+fcb+fc+fb+f+edcb+ed+ec+dcb+dca+da+d+cb+a+1
$$

{\bf Remark.}
The same invariant attack with LZS 444 also appears inside \cite{GenCycleSK19}.
This paper and \cite{GenCycleSK19} can be seen as two radically  different ways to
eventually obtain the same attack.
We believe that the fact that the same attack can be obtained in a many
different ways is very interesting. It is not a bug but rather an feature.
The possibility to obtain the same attack in many distinct ways already occurs inside the framework
of \cite{GenCycleSK19} where we multiply different polynomials (and factorisation is not unique).

\vskip-2pt
\vskip-2pt
\section{Results Achieved and Discussion}
\label{subsec5:96}
\vskip-2pt

Very few non-linear attacks are known and working examples are valuable.
An invariant attack of degree 8 is potentially extremely hard to find or detect, therefore we achieve a stronger attack than before.
It is also substantially stronger in terms of the manipulation of the Boolean function required to make this attack work.
It is easy to see we are closer to an attack which could work for an arbitrary Boolean function.
In order for the attack above to work,
all which we need to do is to make sure that $Z=e$
in $2^{4}$ cases out of $2^{6}$.
These  $2^{4}$ cases correspond to the truth table entries for $Z$ 
such that $(a+b)(c+d)=1$.
All the other $2^{6}-2^{4}$
truth table entries can be arbitrary\footnote{
Or, better, they can be modified to rebalance the Boolean function,
or/and satisfy all sort of desired cryptographic properties such as high non-linearity etc.}.

This shows that a block cipher setup such as LZS 444 can be backdoored
by manipulating only a fraction of entries in a truth table of $Z$.
A recent paper shows that  with another invariant of degree 8 and another LZS wiring this fraction
can be yet smaller, cf. \cite{BackdAnn}.

\vskip-2pt
\vskip-2pt
\subsection{Open Problems}
\label{subsec5:97}
\vskip-2pt

There is still a long way to go in order to actually break T-310.
For KT1 keys we designed an attack of degree 4,
however the Boolean function is quite weak.
then we presented an attack of degree 8 under assumption that
$(Z+e)(a+b)(c+d)=0$ 
which works for stronger Boolean functions
however the long-term key is no longer KT1.
In order to break our cipher with the original Boolean function
we need rather an attack which would work with annihilators
with 3 linear factors like $Z(a+b)c(1+e)=0$,
cf. \cite{MasterPaperT310}.
An example of such an attack can be found in \cite{BackdAnn}
however then again the LZS wiring is not KT1.
Moreover it is possible to see
that current attack constructions (here and in \cite{BackdAnn})
lead to attacks where the annihilating factors involve 2 variables
such as $(a+b)$ which eliminate $F$,
and it is more difficult to construct attacks
with annihilating factors such as $c(1+e)$.
It appears that T-310 disposes of a good security margin
and only weaker cipher setups can currently be attacked. 
Or we need to find better attacks.


%
%

\vskip-2pt
\vskip-2pt
\subsection{The Question of Key Bits}
\label{sec5:Ref}

There exist numerous interesting vulnerabilities in T-310.
For example, the key splits in two parts which are used in a completely different way,
and many other potentially serious vulnerabilities, some of which are listed
in Section 28.2 of \cite{MasterPaperT310}.
In non-linear invariant attacks it iq quite interesting to see that almost
all attacks studied so far do not involve any key bits,
with very few exceptions
(cf.
Section 9.1 in \cite{BackdTut}
We have simply not yet studied the full spectrum of attacks
and attacks which involve key bits could be very useful,
cf. last sub-section in \cite{SlideLCWeakCanAttack310}) and Section 4.11 in \cite{BackdTut}.
In relation to this question,
one anonymous referee for this paper have written the following words:

{\em
[...] when we look carefully at the conditions,
it appears that the first boolean function Z only takes 5 variables [...]
and specifically one bit of the secret key.
Meaning that the secret key only intervenes in a LINEAR way
between the first round and the second round,
I believe other very powerful attacks will appear. [...]
}

\vskip3pt
\noindent
It is an open problem how this property could be exploited in cryptanalysis of T-310.
We should not ignore the fact that the same key bit $L$ is
also used again, just before $Z3$, cf. Fig. \ref{FigComplicationUnit6basic}.
However any variable can potentially be eliminated in an invariant attack.

\vskip-2pt
\vskip-2pt
\section{Conclusion}
\label{sec5}
This 
paper demonstrates a novel invariant hopping technique.
An attack on a pathologically weak cipher setup (with a linear Boolean function!)
is transposed to break another stronger cipher.
In several steps we remove the trivial attacks and keep less trivial ones.
The complexity and algebraic degree for the invariant and the Boolean function,
and the number of variables needed increase progressively.
At the end we obtain an invariant of degree 9 and no invariant of lower degree.

We also show that when the degree of the invariant grows,
in order to make such attacks work one needs to manipulate
only a small fraction of the entries of the truth table of the Boolean function.
This means there exist invariant attacks which are highly compatible
with most known security and non-linearity criteria
on Boolean functions \cite{PeraltaFourMesNL}.
In other words, the attack presented in this paper is not prevented by the
traditional methodology used in block cipher design and analysis.
This in turn means that we need more work on anti-invariant resistance
in block ciphers cf. \cite{BeiCantResNL,GenNonLinRC,FiliolNotVuln}.
Extremely few attacks in symmetric cryptanalysis ever invented
work when the number of rounds is very large.


\end{document}